\documentclass[conference]{IEEEtran}
\IEEEoverridecommandlockouts
\usepackage{cite}
\usepackage{marvosym}
\usepackage{amsmath,amssymb,amsfonts}
\usepackage{algorithmic}
\usepackage{graphicx}
\usepackage{textcomp}
\usepackage{xcolor}
\usepackage{subfigure}
\usepackage{bm}
\usepackage{hyperref}

\def\BibTeX{{\rm B\kern-.05em{\sc i\kern-.025em b}\kern-.08em
    T\kern-.1667em\lower.7ex\hbox{E}\kern-.125emX}}

\begin{document}

\title{iEDA: An Open-Source Intelligent Physical Implementation Toolkit and Library \\
}

\iftrue

\author{
\IEEEauthorblockN{
Xingquan~Li$^{1,11}$\IEEEauthorrefmark{1}, 
Simin~Tao$^{1}$, 
Zengrong~Huang$^{1}$, 
Shijian~Chen$^{1,2,13}$, 
Zhisheng~Zeng$^{1,2,13}$, 
Liwei~Ni$^{1,2,13}$,\\ 
Zhipeng~Huang$^{1}$,
Chunan~Zhuang$^{1}$,
Hongxi~Wu$^{4,3}$,
Weiguo~Li$^{11,3}$,
Xueyan~Zhao$^{2,13,1}$, 
He~Liu$^{5,1}$, 
Shuaiying~Long$^{1}$, \\
Wei~He$^{1}$, 
Bojun~Liu$^{1}$,
Sifeng~Gan$^{1}$, 
Zihao~Yu$^{2,3}$, 
Tong~Liu$^{12,3}$, 
Yuchi~Miao$^{1}$, 
Zhiyuan~Yan$^{2,13}$, 
Hao~Wang$^{6,3}$, \\
Jie~Zhao$^{1}$, 
Yifan~Li$^{7,13,3}$, 
Ruizhi~Liu$^{2,13,1}$,
Xiaoze~Lin$^{1,2,13}$, 
Bo~Yang$^{1,9}$, 
Zhen~Xue$^{1,2,13}$,
Fuxing~Huang$^{4}$,
Zonglin~Yang$^{8,3}$, \\
Zhenggang~Wu$^{1}$, 
Jiangkao~Li$^{11,3}$, 
Yuezuo~Liu$^{6,3}$, 
Ming~Peng$^{6,3}$,  
Yihang~Qiu$^{10,3}$, 
Wenrui~Wu$^{8,3}$, 
Zheqing~Shao$^{6,3}$, \\
Kai~Mo$^{8,3}$, 
Jikang~Liu$^{8,3}$, 
Yuyao~Liang$^{8,3}$, 
Mingzhe~Zhang$^{6,3}$, 
Zhuang~Ma$^{6,3}$, 
Xiang Cong$^{6}$, 
Daxiang Huang$^{1}$,\\
Guojie~Luo$^{5,1}$, 
Huawei~Li$^{2,13,1}$, 
Haihua~Shen$^{13,1}$, 
Mingyu~Chen$^{2,13,1}$, 
Dongbo~Bu$^{14,13,1}$, 
Wenxing~Zhu$^{4,1}$, \\
Ye~Cai$^{8,1}$, 
Xiaoming~Xiong$^{10}$, 
Ying~Jiang$^{9,1}$,
Yi~Heng$^{9,1}$, 
Peng~Zhang$^{1}$, 
Biwei~Xie$^{2,1,\text{\Letter}}$,
~and~Yungang~Bao$^{2,13,3,1,\text{\Letter}}$
}

\vspace{0.4cm}

\IEEEauthorblockA{
$^1$Peng Cheng Laboratory, Shenzhen, China
}
\IEEEauthorblockA{
$^2$State Key Lab of Processors, Institute of Computing Technology, Chinese Academy of Sciences, Beijing, China
}
\IEEEauthorblockA{
$^3$Beijing Institute of Open Source Chip, Beijing, China
}
\IEEEauthorblockA{
$^4$Fuzhou University, Fuzhou, China
}

\IEEEauthorblockA{
$^5$Peking University, Beijing, China
}
\IEEEauthorblockA{
$^6$University of Science and Technology of China, Hefei, China
}
\IEEEauthorblockA{
$^7$Institute of Microelectronics, Chinese Academy of Sciences, Beijing, China
}
\IEEEauthorblockA{
$^8$Shenzhen University, Shenzhen, China
}
\IEEEauthorblockA{
$^9$Sun Yat-sen University, Guangzhou, China
}
\IEEEauthorblockA{
$^{10}$Guangdong University of Technology, Guangzhou, China
}
\IEEEauthorblockA{
$^{11}$Minnan Normal University, Zhangzhou, China
}
\IEEEauthorblockA{
$^{12}$The Hong Kong University of Science and Technology (Guangzhou), Guangzhou, China
}
\IEEEauthorblockA{
$^{13}$University of Chinese Academy of Sciences, Beijing, China
}
\IEEEauthorblockA{
$^{14}$Institute of Computing Technology, Chinese Academy of Sciences, Beijing, China
}

\text{
Email: \IEEEauthorrefmark{1}lixq01@pcl.ac.cn, $^{\text{\Letter}}$xiebiwei@ict.ac.cn,
$^{\text{\Letter}}$baoyg@ict.ac.cn}
}


%

\fi

\maketitle

\begin{abstract}
Open-source EDA shows promising potential in unleashing EDA innovation and lowering the cost of chip design.
This paper presents an open-source EDA project, iEDA, aiming for building a basic infrastructure for EDA technology evolution and closing the industrial-academic gap in the EDA area. iEDA now covers the whole flow of physical design (including Floorplan, Placement, CTS, Routing, Timing Optimization etc.), and part of the analysis tools (Static Timing Analysis and Power Analysis). 
To demonstrate the effectiveness of iEDA, we implement and tape out three chips of different scales (from 700k to 1.5M gates) on different process nodes (110nm and 28nm) with iEDA. iEDA is publicly available from the project home page http://ieda.oscc.cc.


\end{abstract}


\section{Introduction}
\label{sec:Introduction}

The rapid advancement of digital technology has led to a substantial increase in demand for the growth of the integrated circuit (IC) industry. However, Moore's law has reached its limits \cite{lundstrom2022moore}.
To meet the growing needs for computing and storage, we may explore alternative integration technologies, such as 3D chip design or package and chiplets. Moreover, we should strive to improve the quality of chip design, especially in the area of electronic design automation (EDA).
Chip design, EDA tools, and methodologies have been the subject of research for many years. Fortunately, new techniques such as artificial intelligence, hardware acceleration, and collaborative design of chips and EDA have demonstrated their capacity and practical value in EDA. These advancements open up the potential for further optimizing EDA tools and methodologies.





To maximize the benefits of new techniques in EDA, we need an open-source platform accessible to all EDA enthusiasts. There are already several open source physical design platform OpenROAD~\cite{ajayi2019openroad} and tools (e.g., DREAMPlace~\cite{lin2019dreamplace}, CUGR~\cite{liu2020cugr}, Dr.CU 2.0~\cite{li2019dr}, OpenTimer~\cite{huang2015opentimer}) that work well to promote open collaboration and innovation.  
The platform should be automated, well-designed, and compatible while supporting tapeout while offering high performance and scalability, providing comprehensive documentation, and easy-to-learn.
The iEDA project aims to create an open-source EDA platform and build an EDA community. It supports chip implementation from netlist to GDSII for block-level chips and system-on-chips (SoCs) at 28nm process node. The main goal of iEDA is to provide a research platform for enthusiasts to explore chip design, EDA methodologies, algorithms, software development, and more.

\begin{figure}[!t]
    \centering
      \subfigure[]{\label{fig:framework}
      \includegraphics[width=3.7cm]{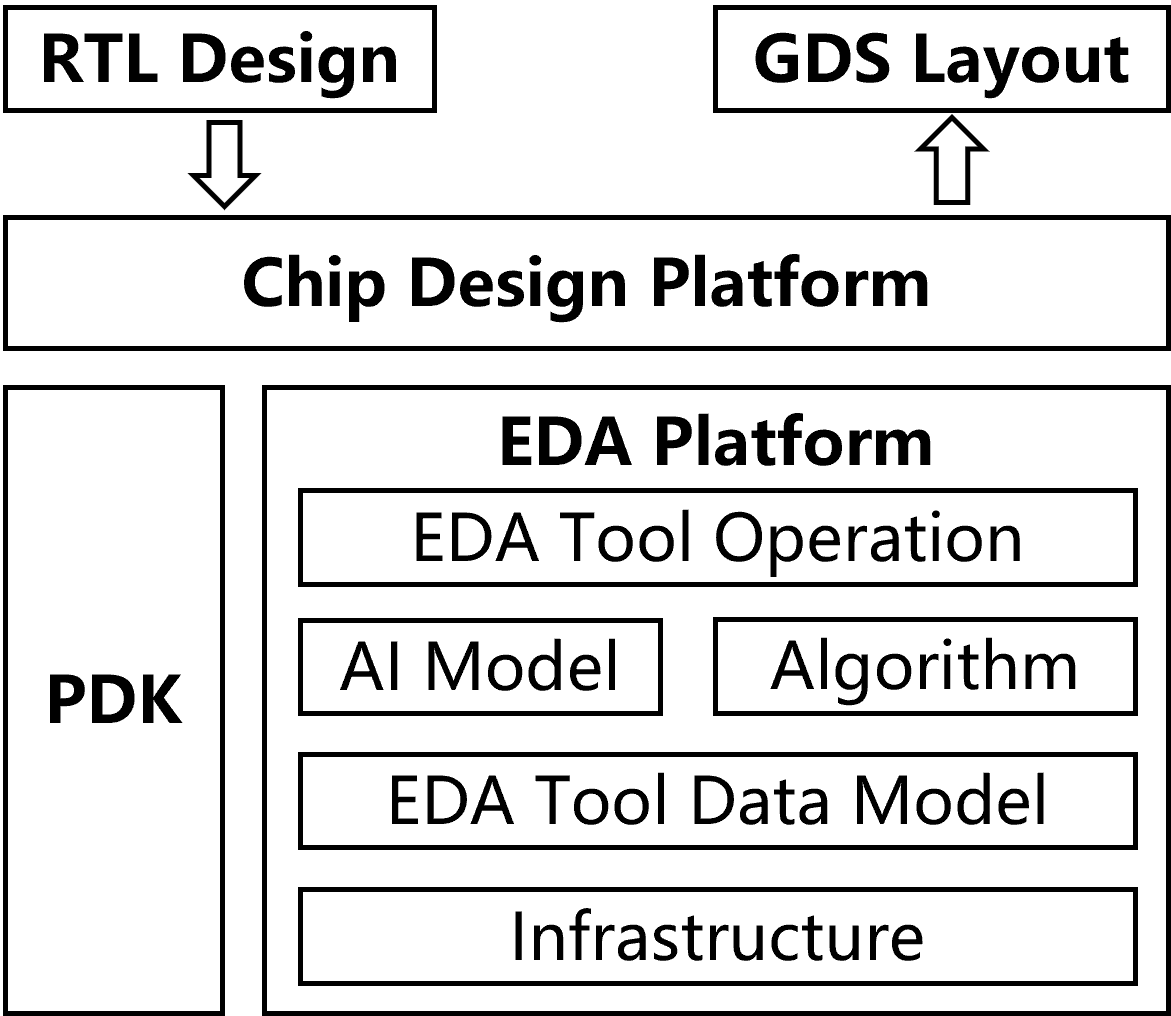}}
      \hspace{0.1cm}
      \subfigure[]{\label{fig:infrastructure}
      \includegraphics[width=4.3cm]{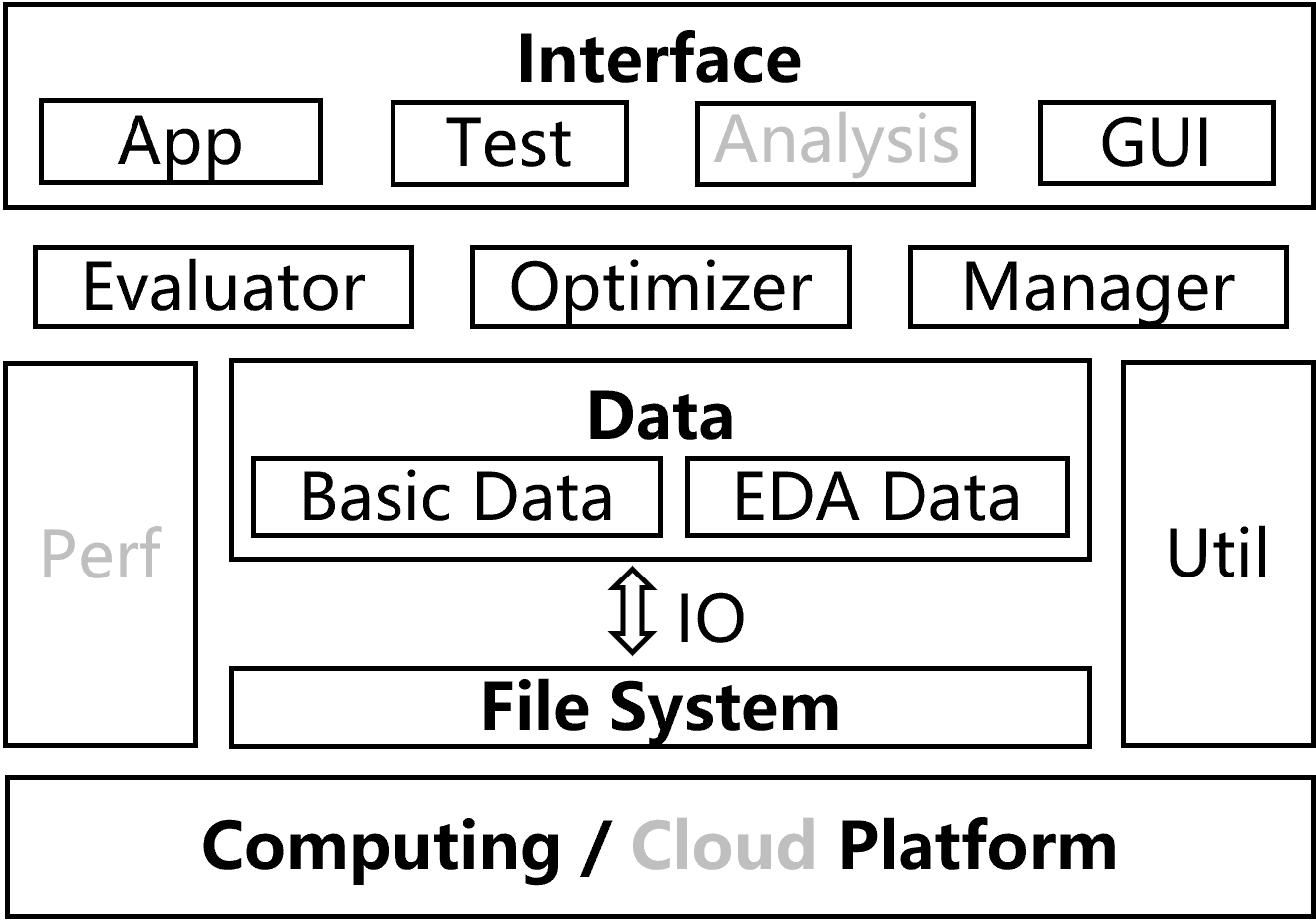}}
    \caption{The iEDA framework and infrastructure.}
    \label{fig:flow}
\end{figure}


\section{iEDA Status}
\label{sec:status}

\subsection{Framework}

The iEDA platform consists of four key components: basic infrastructure, a functional module-level data model, a solver (algorithms and AI models), and function operations as shown in Fig. \ref{fig:framework}. The infrastructure includes the basic data and a series of EDA software kits and frameworks. The data model provides pre-design data services for upper-level EDA functions. The AI models and algorithms are the core solvers that determine the quality and performance of the platform. The operations component is responsible for chip design functions, which include data models and solvers. The iEDA platform, along with a process design kit (PDK), enables the flow of chip design from RTL design to GDS layout.



\subsection{Infrastructure}

The iEDA software infrastructure, as shown in Fig. \ref{fig:infrastructure}, includes a file system with parsers and file managers, a database with builders, writers, and data managers, a platform with evaluators, optimizers, and managers, and an interface with apps, evaluations, and tests, as well as a GUI. This unified infrastructure ensures the robustness and extensibility of iEDA.

\subsection{Tools}

\begin{figure}[!t]
    \centering
    \includegraphics[width=8.8cm]{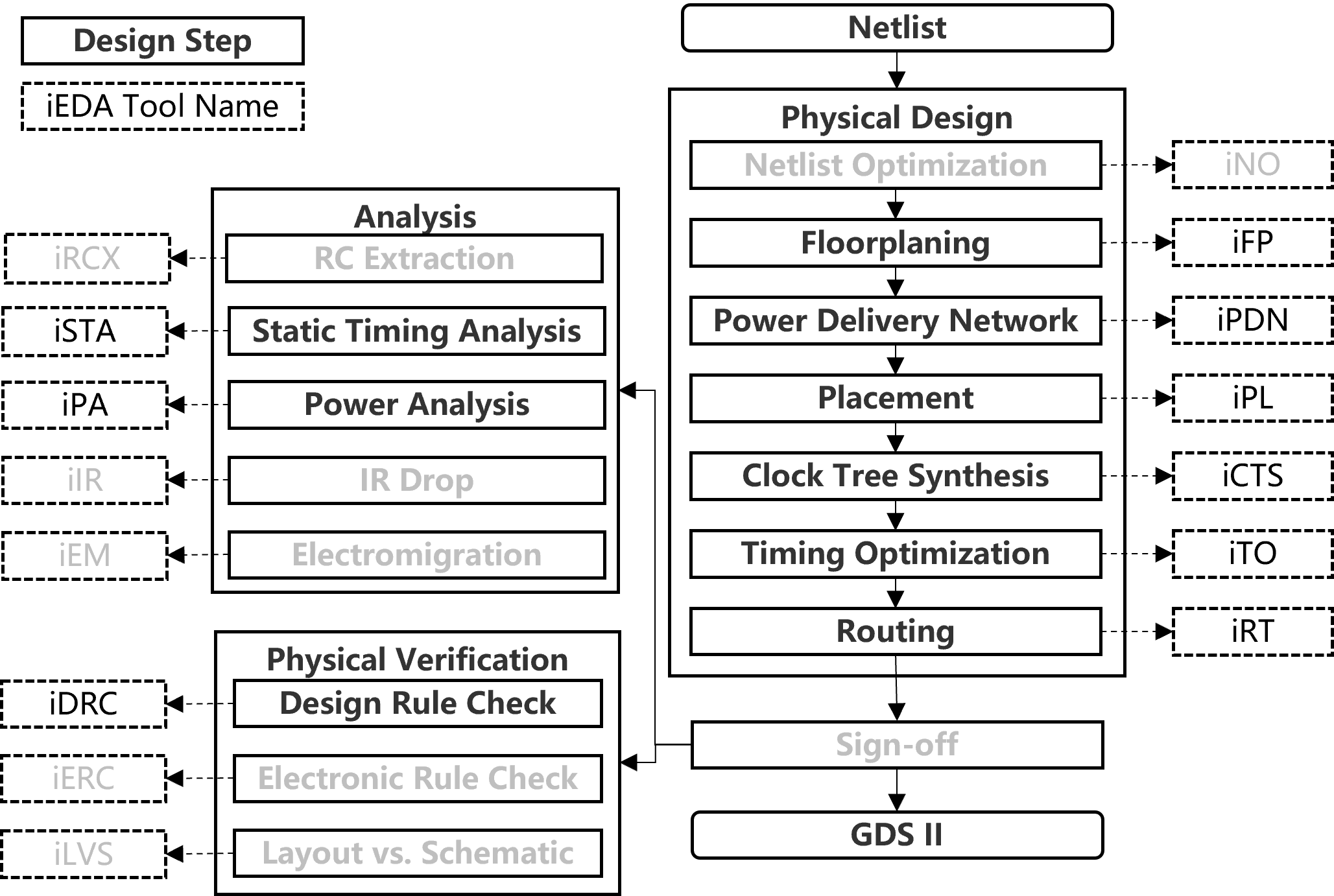}
    \caption{Chip design step and iEDA tools.}
    \label{fig:tool}
\end{figure}

The chip design step and the iEDA tools are shown in Fig. \ref{fig:tool}, each iEDA tool is composed of several low-coupling functional operations, which works by calling a series of different algorithms on designated data models.

\subsubsection{Floorplan and Power Delivery Network}

Floorplanning is the initial stage of physical design. iFP mainly includes the following parts: layout initialization, pre-placement of certain cells, automatic generation of IO locations, automatic placement of macros based on the netlist structure and IO, and generation of the power delivery network.

\subsubsection{Placement} 
Placement mainly ensures the proper coordinate of each cell mapped in the netlist within a designated region, which must comply with design rules and be conducive to routing, timing convergence, and power consumption.
iPL is mainly composed of a global placer, post-global placer, legalizer, detailed placer, filler, and checker. The objective of iPL is to minimize wire-length, timing and congestion.

\subsubsection{Timing Optimization}
%
Timing optimization is aimed at ensuring that the chip design is both functionally correct and meets the design requirements for performance. iTO offers distinct and user-friendly interfaces to fix timing design rule violations, hold time violations, and setup time violations. Our goal is to resolve any timing violations on the chip to enhance its overall performance.

\subsubsection{Clock Tree Synthesis}
The goal of iCTS is to balance skew among flip-flops while optimizing design resource usage, under timing constraints. It has an integrated timing model that offers accurate and efficient computation of timing information, such as delay, capacitance, and slew. The algorithms implemented in iCTS allow for the generation of topologies with various objectives and support the creation of hierarchical clock trees using multiple merging criteria.

\subsubsection{Routing}
Routing entails physical wires of the interconnections among components embedded in a chip. 
The objective of routing is to meet performance and functional requirements while considering factors such as design rule check (DRC) and signal integrity. iRT consists of components such as a resource allocator, topology generator, planar router, layer assigner, guide processor, track assigner, space router, and region manager.

\subsubsection{Static Timing Analysis}

iSTA is a timing analysis tool that provides easy-to-use interfaces for accessing required timing data. Besides of basic timing path analysis, iSTA offers incremental timing propagation, detailed timing reports, advanced delay calculation based on composite current source (CCS) liberty, basic cross-talk delay analysis, and on-chip variation analysis. iSTA also supports parallel processing of load liberty data for faster performance.

\subsubsection{Power Analysis}
iPA is a power analysis tool that supports both basic vectorless analysis without waveform data and vector analysis with VCD/SAIF data. It features fast toggle and static probability propagation, and glitch analysis that considers timing delay.


\subsection{Flow}

\begin{figure}[!t]
    \centering
    \includegraphics[width=8.8cm]{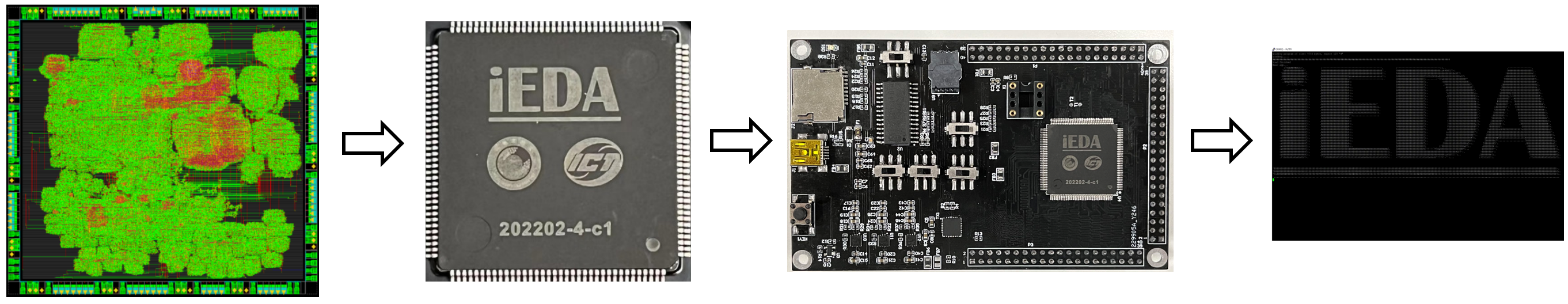}
    \caption{From GDS to computer device.}
    \label{fig:tapeout}
\end{figure}

To verify the functionality and effectiveness of iEDA, we develop iFlow, which is implemented in Python by calling TCL-based or Python-based tool commands. It is worth noting that, since iEDA supports only netlist-to-GDSII, iFlow integrates Yosys/ABC \cite{wolf2013yosys}\cite{brayton2010abc} for synthesis and some commercial tools for signoff analysis and verification.
We implement and tape out three chips with iEDA: one 5-stage RISC-V chip supporting RT-Thread on 110nm; two 11-stage RISC-V chips supporting Linux on 110nm and 28nm in respective.
The first 5-stage RISC-V chip on 110nm has already successfully booted up (Fig. \ref{fig:tapeout}), and the other two chips will return before May.2023.





\section{iEDA Future}
\label{sec:future}

\subsection{Infrastructure}
To better serve the research and development of EDA technology, we will further support more file formats and process nodes, and provide more evaluators and optimizers. In addition, we will try to decompose the tools of iEDA into micro-services and build a cloud-native EDA system.

\subsection{Tools}
The performance of physical design tools will be further improved through carefully designed software architecture and key algorithms, combined with hardware acceleration and AI models, to support larger and more diverse chip designs and scenarios.
We will launch electromigration (EM) and interconnect resistance drop (IR) analysis to complete the analysis toolchain. All analysis tools will share the common database with physical design tools.
Physical verification is the final step in IC design closure and typically involves design rule check (DRC), layout versus schematic (LVS), and electrical rule check (ERC). In the upcoming version of iEDA, the development of these tools is being planned. 


\subsection{Flow}


Chip design flow encompasses multiple steps and involves numerous parameters. Furthermore, the requirements for chip design and process node are highly diverse, making it challenging to adapt to various design objectives and constraints within a specific chip design script. To overcome this challenge and achieve an efficient flow with the right parameter configuration, our plan is to extract the majority of parameters from the EDA tools used in the design flow. Additionally, we aim to optimize the chip design objective by conducting a thorough study of the design space exploration

\subsection{AiEDA}
\begin{figure}[!t]
    \centering
    \includegraphics[width=8.9cm]{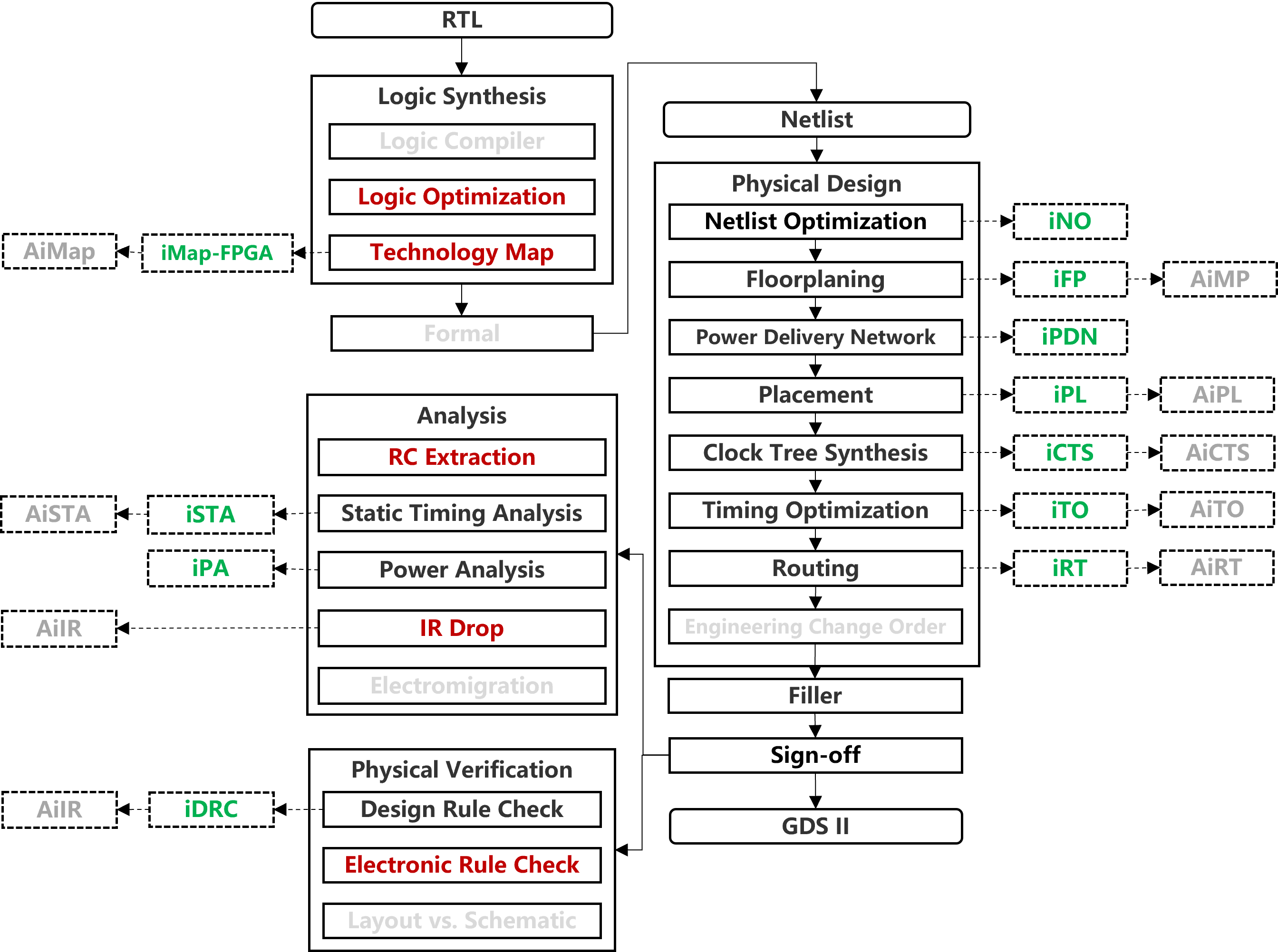}
    \caption{iEDA tools with AI models.}
    \label{fig:tool-ai}
\end{figure}

%
In the era of intelligence, it has gradually become a trend to use artificial intelligence (AI) technology to liberate engineers from heavy repetitive work. This can also further improve the chip performance and production efficiency. With the powerful  application of AI in many  engineering fields,  we have reason to believe that AI can bring significant value for EDA. However, due to the deficiency of AI infrastructure in EDA, such as open-source EDA toolkit and package, accessible labeled data set and system, and valuable AI model for EDA, the existing AI works in EDA lack practical implementation possibilities. To maximize the value of AI for EDA, we will build some AI infrastructures in the next iEDA version. Futhermore, AI researcher and developer can focus on valuable AI model for EDA, and then we will obtain a better chip design solution which is composed of EDA tool and AI model as shown in Fig. \ref{fig:tool-ai}.

\section*{Acknowledgment}
This work is supported in part by the Major Key Project of PCL (No. PCL2021A08), the Strategic Priority Research Program of Chinese Academy of Sciences (No. XDC05030400), the Strategic Priority Research Program of Chinese Academy of Sciences (No. XDA0320300), the National Key R\&D Program of China (No. 2022YFB4500403), National Natural Science Foundation of China (No. 62090024), the National Natural Science Foundation of China (No. 62090021),  the National Natural Science Foundation of China (No. 62174033),  the National Natural Science Foundation of China (No. 61907024).

\bibliographystyle{unsrt}

\end{document}